\documentclass[useAMS,usenatbib]{mn2e}
\usepackage{times}
\usepackage{graphicx}

\title[A resonance model with magnetic connection]
  {A resonance model with magnetic connection for 3:2 HFQPO pairs in black hole binaries}
\author[Chang-Yin Huang et al.]
  {Chang-Yin Huang,$^1$
  Zhao-Ming Gan,$^1$ Jiu-Zhou Wang,$^1$ Ding-Xiong Wang$^1$\thanks{E-mail: dxwang@mail.hust.edu.cn} \\
  $^1$School of Physics, Huazhong University of Science and Technology, Wuhan 430074, China\\}
\date{24 December 2009}

\pagerange{\pageref{firstpage}--\pageref{lastpage}} \pubyear{0000}

\def\LaTeX{L\kern-.36em\raise.3ex\hbox{a}\kern-.15em
    T\kern-.1667em\lower.7ex\hbox{E}\kern-.125emX}

\begin{document}

\label{firstpage}

\maketitle

\begin{abstract}
We apply epicyclic resonances to the magnetic connection (MC) of a
black hole (BH) with a relativistic accretion disc, interpreting the
high frequency quasi-periodic oscillations (HFQPOs) with 3:2 pairs
observed in three BH X-ray binaries. It turns out that the 3:2 HFQPO
pairs are associated with the steep power-law states, and the severe
damping can be overcome by transferring energy and angular momentum
from a spinning BH to the inner disc in the MC process.
\end{abstract}

\begin{keywords}
accretion, accretion discs -- black hole physics -- magnetic fields
-- stars: individual (GRO J1655-40, GRS 1915+105, XTE J1550-564) --
stars: oscillations -- X-rays: stars
\end{keywords}

\section{Introduction}

As is well known, the high frequency Quasi-Periodic Oscillations
(HFQPOs) have been observed in several X-ray Binaries. As argued by
\citet{b30,b31}, HFQPOs in X-ray binaries probably originate from
the inner edge of an accretion disc around a black hole (BH) of
stellar-mass, since millisecond is the natural timescale for
accretion process in these regions. Although a number of models have
been proposed to explain the HFQPOs in X-ray binaries, no consensus
has been reached on their physical origin (see a review by
\citealt{b13}, hereafter MR06). The 3:2 HFQPO pairs have been
observed in a few BH binaries, i.e., GRO J1655-40 (450, 300Hz;
\citealt{b22}; \citealt{b27}; \citealt{b23}), XTE J1550-564 (276,
184Hz; \citealt{b15}; \citealt{b23}) and GRS 1915+105 (168, 113Hz;
MR06; \citealt{b25}, hereafter RM06).

It has been pointed out that HFQPOs are generally associated with
the steep power-law (SPL) state in BH X-ray binaries. Although the
3:2 HFQPO pairs could be interpreted in some epicyclic resonance
models \citep{b1,b2,b10,b29}, there remain serious uncertainties as
to whether epicyclic resonance could overcome the severe damping
forces and emit X-rays with sufficient amplitude and coherence to
produce the HFQPOs (e.g., see a review in MR06).

Very recently, \citet{b6} proposed a model of magnetically induced
disc-corona to interpret the spectrum states in some BH binaries, in
which the magnetic connection (MC) of a BH with its surrounding
accretion disc can be produced naturally based on several minimal
assumptions. It is found that the spectrum states in the BH binaries
depend on accretion rate and BH spin, and henceforth this model is
referred to as the MC model.

In this paper we intend to combine the resonance model with the MC
model to fit the association of the 3:2 HFQPO pairs with the SPL
state in the BH binaries. This paper is organized as follows. In
Section 2 we discuss the association of HFQPOs with the SPL state
based on the MC model. In Section 3 we discuss the input of the
electromagnetic energy onto the resonance modes to overcome the
severe damping by transferring energy from the BH to the inner disc
via the MC process. Finally, in section 4 we discuss some issues
related to our model. Throughout this paper the geometric units
$G=c=1$ are used.

\section{ASSOCIATION OF HFQPOS WITH SPL STATES}

G09 resolved the dynamic equations of standard thin disc model,
combining the MC effects in a typical disc-corona scenario, and the
main points are summarized as follows. The interior viscous stress
is assumed to be proportional to gas pressure, which is comparable
to magnetic pressure. The interior viscous process is dominantly
governed by tangled small-scale magnetic fields. A fraction of the
viscously dissipated energy is released in the disc, emitting
eventually as blackbody radiation and supplying seed photons for
Comptonization of corona, and the rest viscously dissipated energy
is converted into energy in the corona, heating corona and
maintaining its relativistic temperature. It is assumed that the
large-scale magnetic field is generated from the small-scale
magnetic field, between which the relation is given as follows
\citep{b12},

\begin{equation}
   B_{\textup{\scriptsize p}}\sim(h/r)\cdot B_{\textup{\scriptsize D}}.
\end{equation}

In equation (1) $B_{\textup{\scriptsize p}}$ is the poloidal
component of the large-scale magnetic field anchored on the disc,
and $h$ is the half height of the disc. The rotational energy of a
BH can be extracted via the MC process, which contributes to the
energy dissipation of the disc-corona system, and the mapping
relation between the BH horizon and the inner disc can be determined
based on the conservation of magnetic flux as follows,

\begin{equation}
  B_{\textup{\scriptsize p}}\cdot 2\pi r \sqrt{\frac{A}{D}}\cdot
   \textup{d}r|_{\theta=\pi/2}
   =-B_{\textup{\scriptsize H}} \cdot 2\pi r_{\textup{\scriptsize H}} \cdot 2M\sin\theta\textup{d}\theta|_{r=r_{\textup{\tiny
   H}}},
\end{equation}
where the left and right sides represent the magnetic fluxes
threading the disc and the BH horizon, respectively. The quantities
$M$, $r_{\textup{\scriptsize H}}$ and $B_{\textup{\scriptsize H}}$
are the BH mass, the horizon radius and the magnetic field at the
horizon, respectively, and $\theta$ is the angular coordinate of the
horizon. The quantities $A$ and $D$ in equation (2) are relativistic
correction factors in Kerr metric, and they read

\begin{equation}
   A=1+a_{*}^{2}/\tilde{r}^{2}+2a_{*}^{2}/\tilde{r}^{3},\ \
   D=1-2/\tilde{r}+a_{*}^{2}/\tilde{r}^{2}.
\end{equation}
where $a_{*}\equiv J/M^{2}$ is the BH spin defined in terms of the
BH mass $M$ and BH angular momentum $J$, and $\tilde{r}\equiv r/M$
is dimensionless disc radius.

In our model we assume that the magnetic field is uniform at the
horizon, and it is related to accretion rate
$\dot{M}_{\textup{\scriptsize D}}$ based on the equilibrium between
magnetic pressure and the ram pressure given by \citet{b20},

\begin{equation}
   B_{\textup{\scriptsize H}}=\sqrt{2\dot{M}_{\textup{\scriptsize D}}/r_{\textup{\scriptsize
   H}}^{2}},
\end{equation}

As argued in G09, the energy dissipation in the MC process can be
derived by resolving relativistic conservation equations of energy
and angular momentum for a disc with the MC. The X-ray spectra of
the BH binaries can be simulated by resolving the magnetic
disc-corona system with Monte-Carlo method.

According to the epicyclic resonance model \citep{b3,b29}, the 3:2
HFQPO pairs are fitted by the radial and vertical resonance
frequencies as follows,

\begin{equation}
   \nu_{r}=\nu_{\phi}(1-6\tilde{r}^{-1}+8a_{*}\tilde{r}^{-3/2}-3a_{*}^{2}\tilde{r}^{-2})^{1/2},
\end{equation}

\begin{equation}
   \nu_{\theta}=\nu_{\phi}(1-4a_{*}\tilde{r}^{-3/2}+3a_{*}^{2}\tilde{r}^{-2})^{1/2},
\end{equation}

\begin{equation}
   \nu_{\theta}/\nu_{r}=3/2,
\end{equation}
where
\begin{equation}
   \nu_{\phi}=\frac{\nu_{0}}{2\pi m_{\textup{\scriptsize
   BH}}(\tilde{r}^{3/2}+a_{*})}, \\ \nu_{0}\equiv
   M_{\odot}^{-1}=2.03\times 10^{5}{\textup {Hz}}.
\end{equation}

In equations (5)---(8) $\nu_{\phi}$ is the Keplerian frequency,
$\nu_{r}$ and $\nu_{\theta}$ are respectively the radial and
vertical resonance frequencies of the general relativistic disc, and
$m_{\textup{\scriptsize BH}}$ is the BH mass in terms of the solar
mass.

By using equations (5)---(8) we obtain the resonance radii and the
BH spins of the three BH binaries for the given BH mass in fitting
the 3:2 HFQPO pairs. In addition, we can fit the SPL states of these
sources by combining the resonance model with the MC model given in
G09, and the concerned observational quantities and the fitting
parameters are listed in Tables 1 and 2, respectively.

\begin{table*}
 \centering
 \begin{minipage}{110mm}
  \caption{The observational quantities of the BH binaries with the 3:2 HFQPO pairs.}
  \begin{tabular}{@{}lccccc@{}}
  \hline
   Source & $m_{\textup{\scriptsize BH}}$$^{a}$ & {\it D} (kpc)$^{a}$ & {\it i} (degree) & $\nu_{\textup{\scriptsize
   QPO}}$(Hz)$^{a}$ & $L_{\textup{\scriptsize X,SPL}}$($L_{\textup{\scriptsize Edd}}$)$^{b}$ \\
 \hline
 GRS 1915+105 & 10 -- 18 & 11 -- 12 & $70\pm2$$^{c}$ & 168,113 & 1.1 \\
 XTE J1550-564 & 8.4 -- 10.8 & $5.3\pm2.3$ & 74$^{d}$ & 276,184 & 0.5 \\
 GRO J1655-40 & 6.0 -- 6.6 & $3.2\pm0.2$ & 70$^{e}$ & 450,300 & 0.1 \\
\hline
\end{tabular}
\newline
\newline $^{a}$MR06; $^{b}$\citet{b5}; $^{c}$\citet{b7}; $^{d}$\citet{b21}; $^{e}$\citet{b4}
\end{minipage}
\end{table*}

\begin{table*}
 \centering
 \begin{minipage}{155mm}
  \caption{The parameters in fitting the SPL state of the BH binaries with the 3:2 HFQPO pairs.}
  \begin{tabular}{@{}lcccccccccc@{}}
  \hline
   Source & $m_{\textup{\scriptsize BH}}$ & $a_{*}$ & $r_{\textup{\scriptsize 32}}/r_{\textup{\scriptsize ms}}$ &
    $r_{\textup{\scriptsize out}}/r_{\textup{\scriptsize ms}}$  & $\dot{m}$
     & $\alpha$ & Photon Index & $L_{\textup{\scriptsize X}}$($L_{\textup{\scriptsize Edd}}$)&
     $F_{\textup{\scriptsize disc}}/F_{\textup{\scriptsize total}}$ & $F_{\textup{\scriptsize PL}}/F_{\textup{\scriptsize total}}$\\
 \hline
 GRS 1915+105 & 10 & 0.685 & 1.931 & 14.255 & 0.250 & 0.300& 2.875 & 0.460 & 0.274 & 0.722\\
 & 18 & 0.994 & 2.895 & 18.046 & 0.088 & 0.390 & 2.466 & 0.927 & 0.383& 0.612\\
 \hline
 XTE J1550-564 & 8.4 & 0.888 & 2.114 & 19.670 & 0.130 & 0.300 & 2.348& 0.442 & 0.234 & 0.757\\
 & 10.8 & 0.990 & 2.741 & 16.913 & 0.100 & 0.391 & 2.615& 0.897 & 0.420 & 0.576 \\
 \hline
 GRO J1655-40 &  6.0 & 0.955 & 2.332 & 19.463 & 0.110 & 0.300 & 2.585& 0.549 & 0.347 & 0.646\\
 &  6.6 & 0.989 & 2.712 & 19.380 & 0.090 & 0.335 & 2.592& 0.754 & 0.400 & 0.593 \\
\hline
\end{tabular}
\end{minipage}
\end{table*}

\begin{table*}
 \centering
 \begin{minipage}{95mm}
  \caption{Comparison of transporting electromagnetic energy in BH magnetosphere with DC circuit.}
  \begin{tabular}{@{}lcc@{}}
  \hline
   Sources & BH magnetosphere & Steady DC circuit \\
 \hline
 Extracted Energy & Spinning energy of a BH & Non-electromagnetic energy \\
 Load & Inner disc & Resistance \\
 Transport Direction & From BH to inner disc & From battery to resistance  \\
 Energy Conversion & Magnetic reconnection & Dissipation in Joule
 heating \\
 Electron Accelerating & Induced Electric field & Steady Electric
 field\\
 Radiation & Inverse-Compton & None \\
 Energy Transport & Poynting flux & Poynting flux\\
 \hline
\end{tabular}
\end{minipage}
\end{table*}

The quantities $m_{\textup{\scriptsize BH}}$, $D$ and $i$ in Table 1
are the BH mass in terms of the solar mass, the distance estimated
of each source and the orbital inclination angle, respectively. The
values of $r_{\textup{\scriptsize 32}}/r_{\textup{\scriptsize ms}}$
in Table 2 are the resonance radii in terms of the radius of the
innermost stable circular orbit (ISCO), being obtained by invoking
$\nu_{\theta}/\nu_{r}=3/2$. The parameters $\dot{m}$ and $\alpha$
are the accretion rate in terms of Eddington accretion rate and the
viscous efficiency, respectively.

\begin{figure}
\vspace{0.6cm}
\begin{center}
{\includegraphics[width=8.5cm]{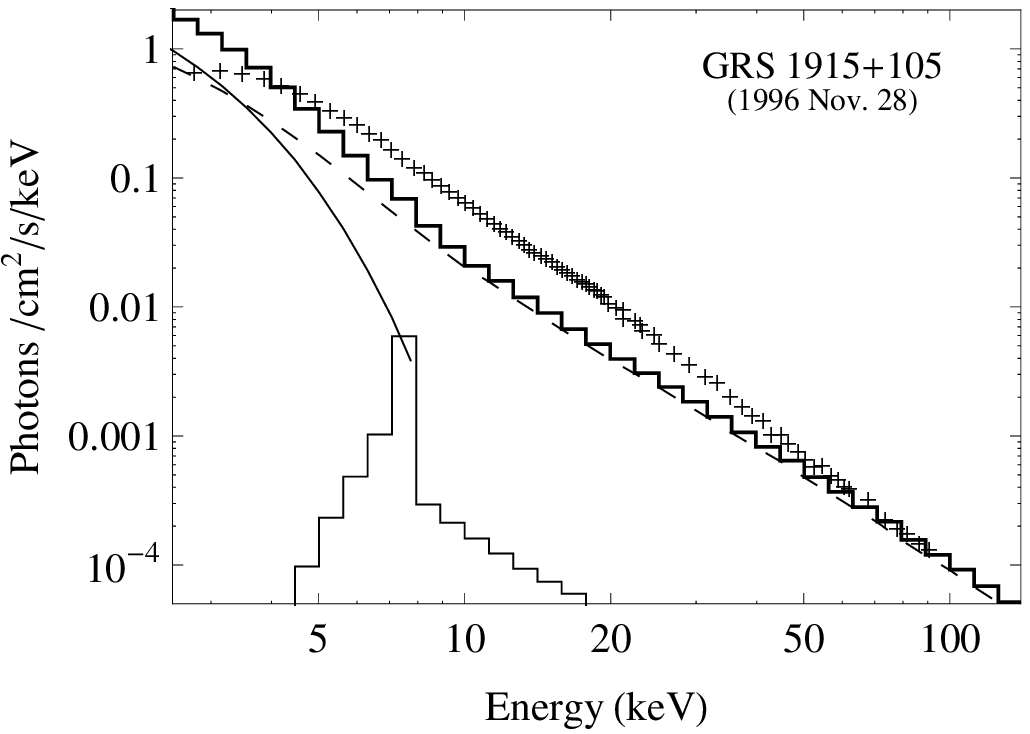}}
{\includegraphics[width=8.5cm]{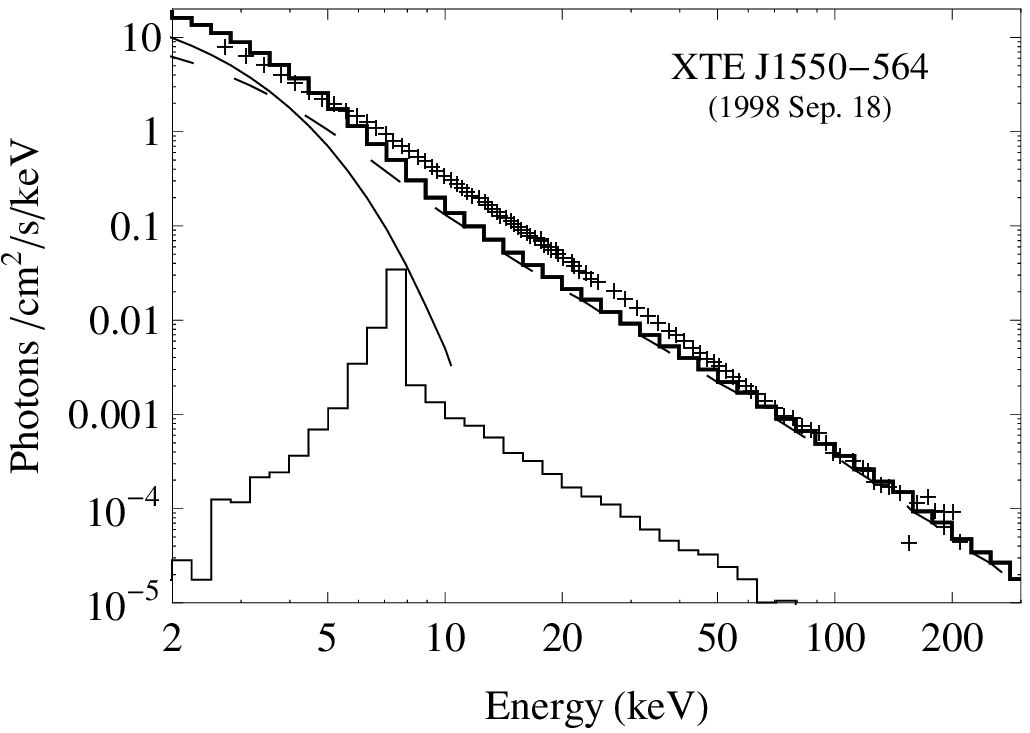}}
{\includegraphics[width=8.5cm]{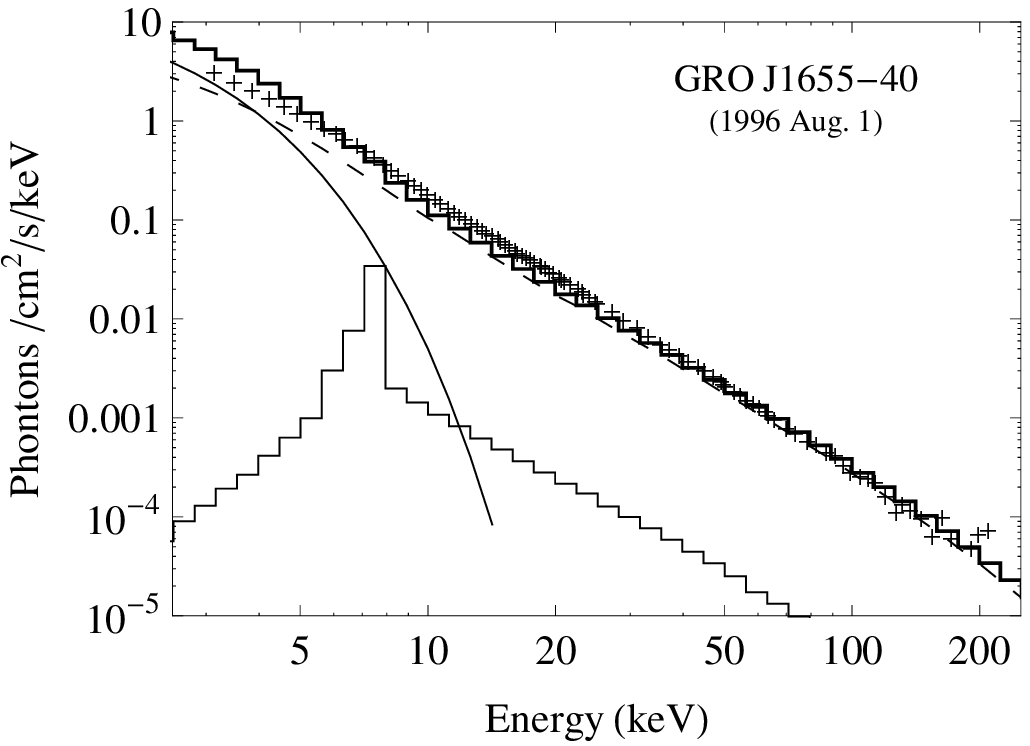}}
 \caption{Emerged spectrum of the three sources in the SPL state:
 (a) GRS 1915+105 (1996 Nov. 28) fitted with $m_{\textup{\scriptsize BH}}$=18, $\dot{m}$=0.088 and
 $\alpha$=0.39 in the upper panel, (b) XTE J1550-564 (1998 Sep. 18)
 fitted with $m_{\textup{\scriptsize BH}}$=10.8, $\dot{m}$=0.1 and $\alpha$=0.391 in the middle panel,
 (c) GRO J1655-40 (1996 Aug.1) fitted with $m_{\textup{\scriptsize BH}}$=6.6, $\dot{m}$=0.09,
  $\alpha$=0.335 in the bottom panel. The total emissive spectrum and its thermal,
  comptonized and reflective components are plotted in thick-zigzag, thin-solid,
  dashed and thin-zigzag lines, respectively. The observation data (without error bars) of
  GRS 1915+105 are taken from Fig. 4.13 of MR06, and those of GRO J1655-40 and XTE J1550-564
  taken respectively from Figs. 1 and 2 of \citet{b24}.}
\end{center}
\end{figure}

Inspecting Table 2, we find that the values of the concerned
parameters are in accordance with the definition of the SPL state
given in RM06, and some issues in the fittings are given as follows.

(i) As shown in Table 2, the resonance radius
$r_{\textup{\scriptsize 32}}$ is less than the radius
$r_{\textup{\scriptsize out}}$ , i.e., $r_{\textup{\scriptsize 32}}$
is located in disc region covered by corona, and the radiation from
the resonance mode is influenced directly by the corona.

(ii) We take the upper limit to the BH mass with the lower limit to
the distance from the observers to fit the spectra of the three
sources, since the bigger BH mass and the less distance give rise to
the stronger photon flux in the fittings based on the MC model.

(iii) As shown in Table 2, the power-law component dominates
significantly over the disc component for each source. The
adjustable parameters are only accretion rate and viscous
coefficient $\alpha$ for the given 3:2 HFQPO pairs. It is found that
the soft ($<\sim$10 keV) and hard ($>\sim$10 keV) parts of the
power-law component in the radiation increase and decrease with the
increasing accretion rate, respectively, while the corresponding
fractions increase and decrease with the decreasing viscous
coefficient $\alpha$. Thus, as shown in Table 2, we fit the bigger
photon index in the SPL state by invoking greater accretion rate.
The high values of $\alpha$ given in Table 2 arise probably from the
differential rotation strengthened due to the transfer of angular
momentum from the spinning BH to the inner disc in the MC process.

(iv) The higher BH spins are required in fitting the 3:2 HFQPO pairs
spin based on equations (5)---(8). On the other hand, according to
the MC model, the higher spin corresponds to the lower hardness of
the power-law component in the radiation, giving rise to the steeper
photon index. Thus we have the association of the 3:2 HFQPO pairs
with the SPL state in these BH binaries.

Based on the MC model we have the emerged spectra of SPL states of
the three sources as shown in Figure 1.

\begin{figure}
\begin{center}
{\includegraphics[width=8cm]{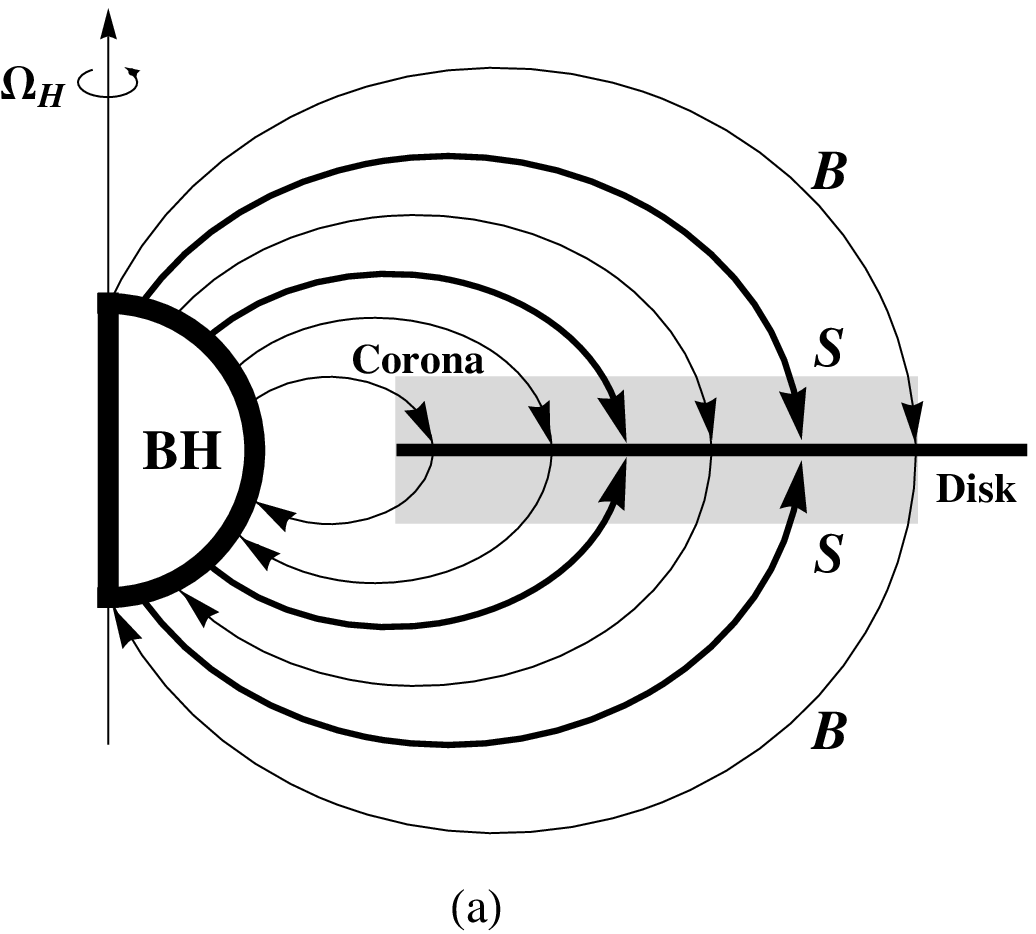}
\includegraphics[width=7.5cm]{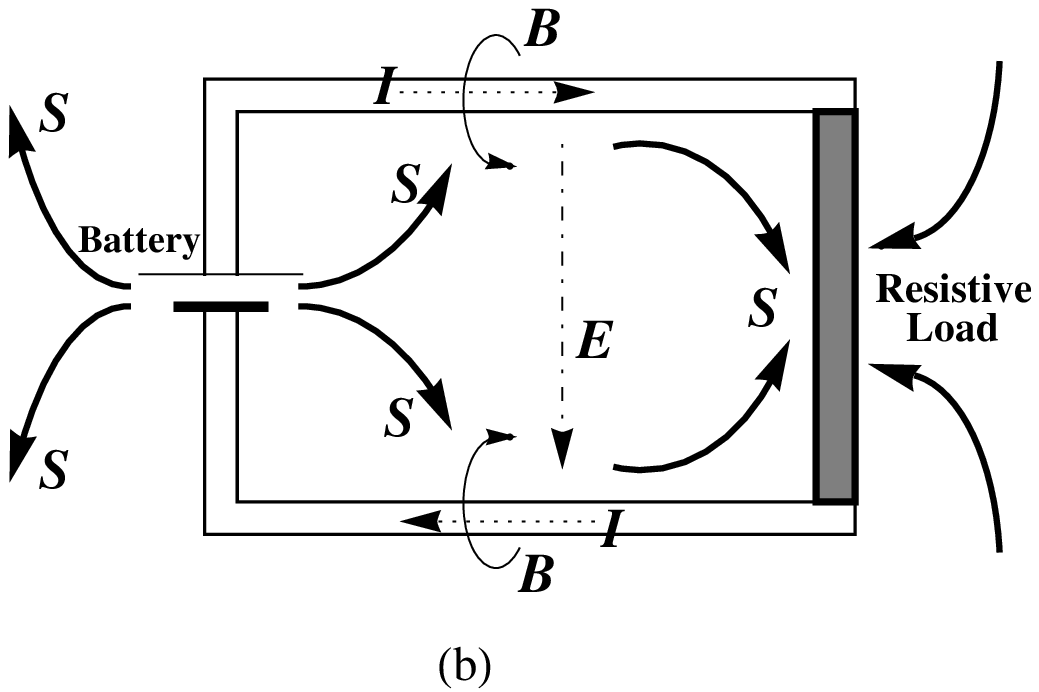}}
 \caption{(a) A magnetosphere with closed field lines connecting a
 spinning BH with its surrounding disc is shown in the upper panel,
 where the thin and thick arrows represent the closed magnetic field
 and Poynting energy flux, respectively. (b) A steady circuit is shown
 in the lower panel, where the thin and dot-dashed arrows represent
 magnetic and electric fields, respectively. The thick solid and thin
 dotted arrows represent Poynting energy flux and the current in the
 circuit, respectively. }
\end{center}
\end{figure}

\section{ENERGY INPUT ONTO RESONANCE MODES}

In order to understand the input of the electromagnetic energy onto
the resonance modes, we have an analogy of the BH magnetosphere to
an electric circuit as shown in Figure 2. In both cases we find that
the electromagnetic energy is transported to the load as Poynting
flux. As argued by \citet{b32} the energy is extracted magnetically
from a fast-spinning BH, and transferred to the inner disc via the
closed field lines in the MC process as shown in the upper panel of
Figure 2. This scenario is very similar to the energy transfer from
a battery to the resistive load in a circuit as shown in the lower
panel of Figure 2. In the circuit electrons are accelerated by the
electric field, resulting in the dissipation of the kinetic energy
in Joule heating. In the MC process the electromagnetic energy
transferred into the inner disc could be converted into the kinetic
energy of electrons by the induced electric field due to the
magnetic reconnection. In this way the severe damping in resonance
modes could be avoided.

On the other hand, the magnetic reconnection is more likely to occur
due to the resonance of the plasmoids in the accretion flow, and the
electrons in the corona could be accelerated to upscatter soft
photons to hard photons in inverse-Compton radiation. This scenario
is helpful to understand the association of the HFQPO with
non-thermal radiation observed in the SPL state of the BH binaries.
A rough comparison of the two processes of transporting
electromagnetic energy is given in Table 3.

\begin{figure}
\vspace{0.6cm}
\begin{center}
{\includegraphics[width=5cm]{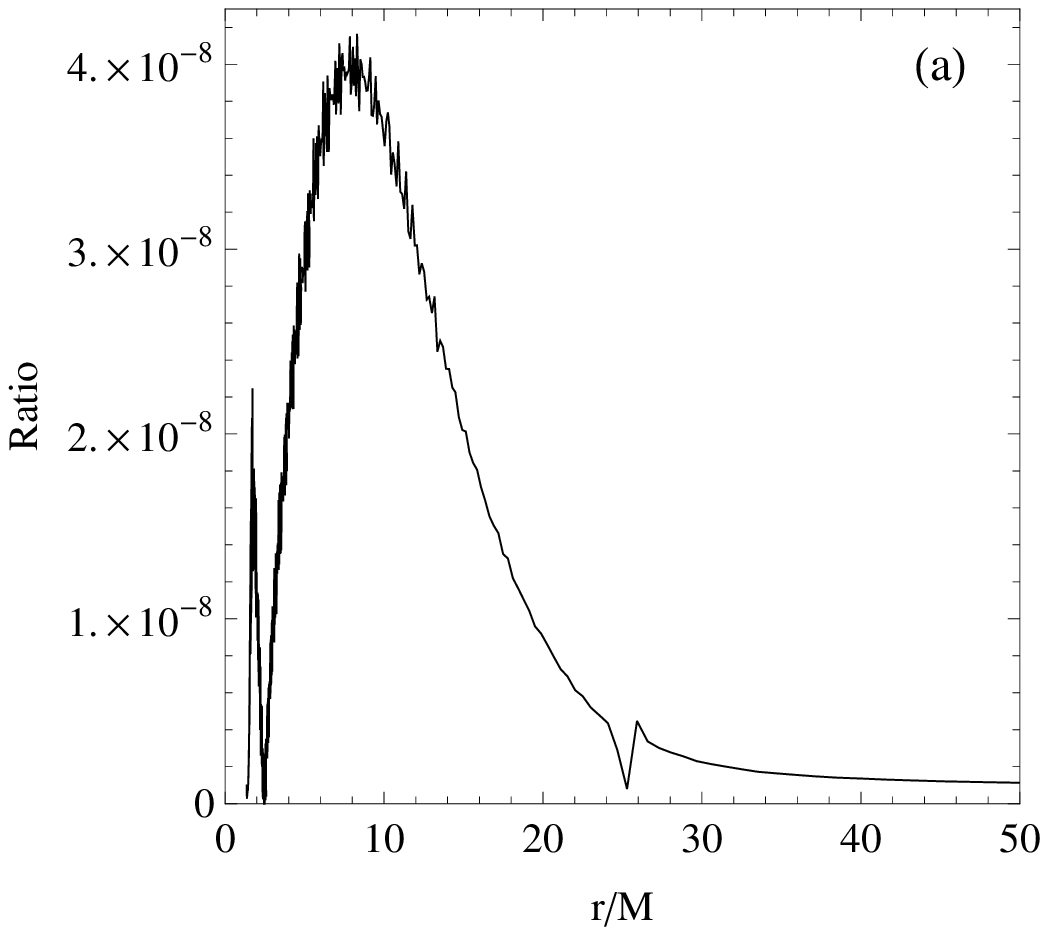}
\includegraphics[width=5cm]{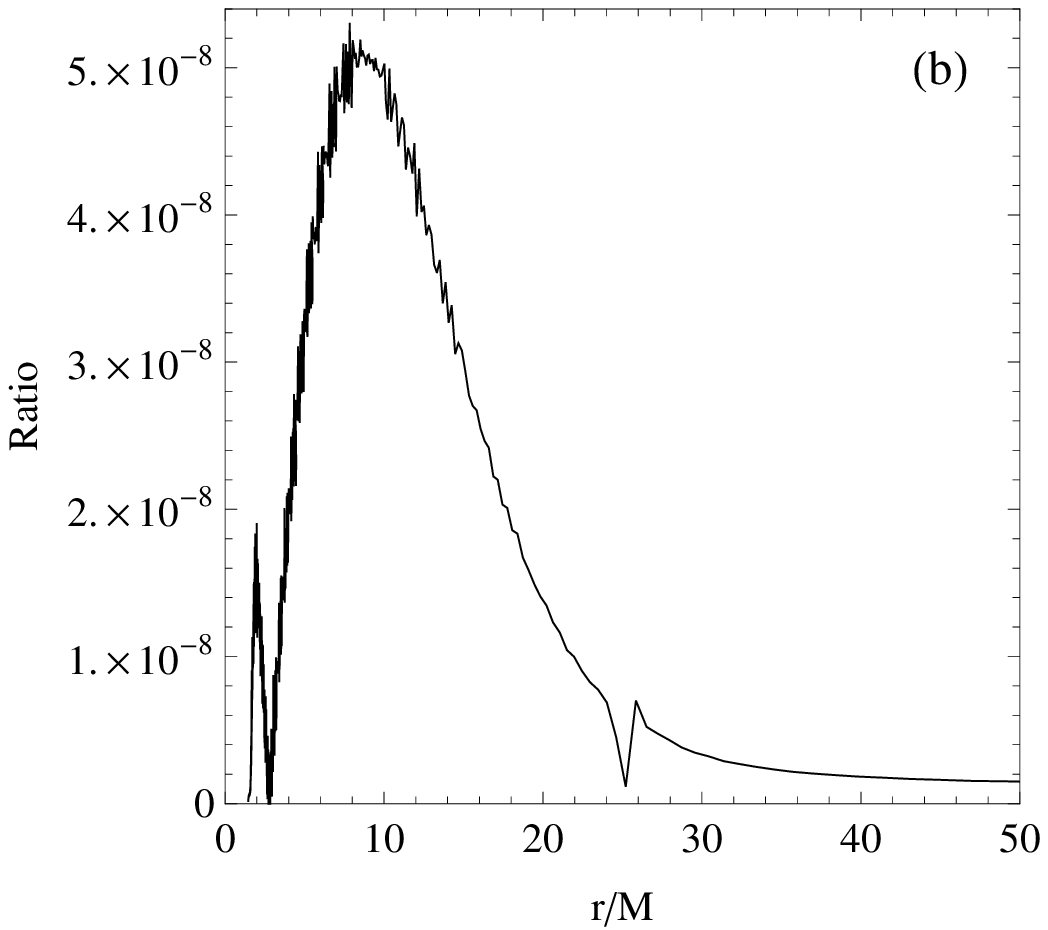}
\includegraphics[width=5cm]{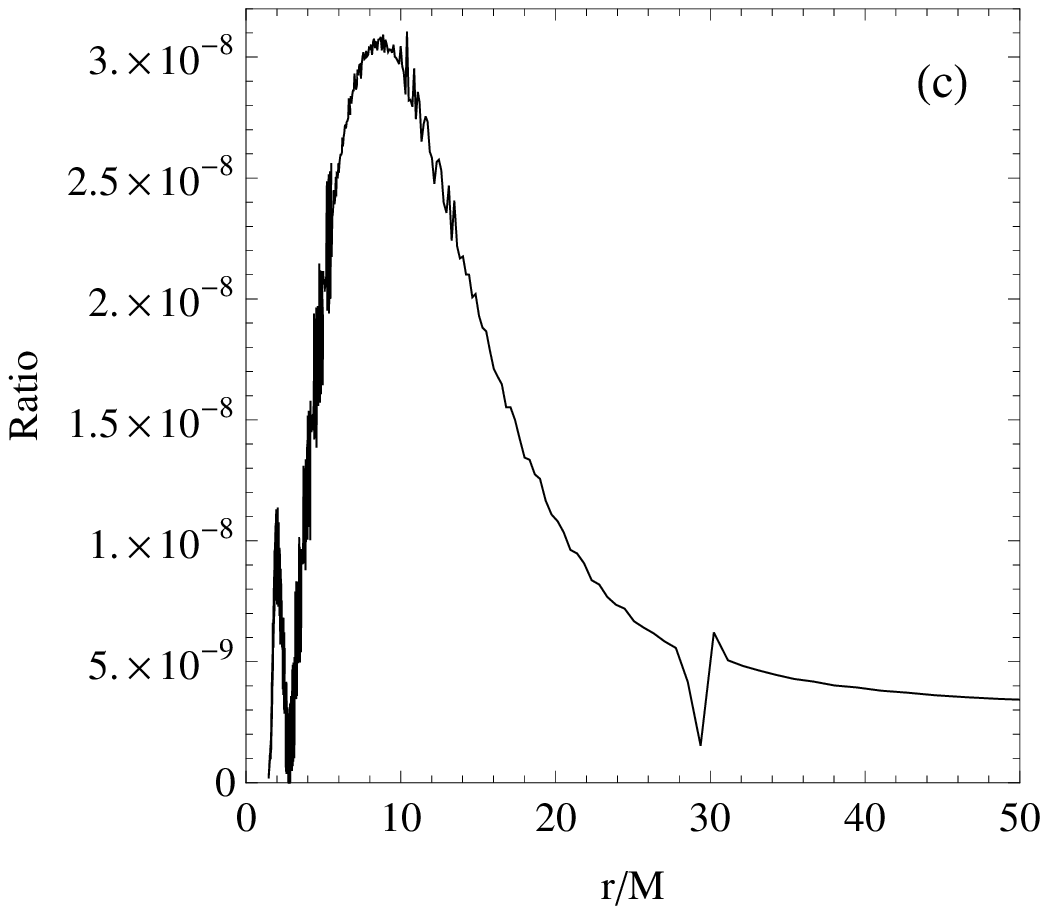}}
 \caption{The ratio of $|\nabla(B^{2}/8\pi)|$ to $\rho\Omega^{2}r$ versus the disc radius in the scenario for
 the 3:2 HFQPO pairs. (a) GRS 1915+105 with $m_{\textup{\scriptsize BH}}$=18, $\dot{m}$=0.088 and
 $\alpha$=0.39; (b) XTE J1550-564 with $m_{\textup{\scriptsize BH}}$=10.8, $\dot{m}$=0.1 and
 $\alpha$=0.391 and (c) GRO J1655-40 with $m_{\textup{\scriptsize BH}}$=6.6, $\dot{m}$=0.09,
  $\alpha$=0.335.}
\end{center}
\end{figure}

\section{DISCUSSION}

In this paper, we combine the epicyclic resonances in a relativistic
accretion disc with the MC to interpret the 3:2 HFQPOs pairs in the
BH X-ray binaries. It turns out that the 3:2 HFQPO pairs are
associated with the SPL state, and the severe damping can be
overcome by transferring energy from a spinning BH to the inner disc
via the MC process. Some issues related to this scenario are
discussed as follows.

\subsection{Weak Magnetic Field Assumption}

\citet{b11} discussed the condition of quasi-steady state in the MC
of a spinning BH with an accretion disc, and he assumed that the
magnetic field is weak so that its influence on the dynamics of
particles in a thin Keplerian disc can be negligible. The ``magnetic
field assumption'' is given by

\begin{equation}
   |\nabla(B^{2}/8\pi)|\ll\rho|g|,
\end{equation}
where $\rho$ is the mass density of the disc and $g$ is the
gravitational acceleration produced by the BH.

It is easy to check that the assumption given by equation (9) is
valid in our scenario for the 3:2 HFQPO pairs, so that the epicyclic
resonance radii can be determined in the context of general
relativity. Replacing the gravitational acceleration $g$ in equation
(9) by $\Omega^{2}r$, we have

\begin{equation}
   Ratio=\frac{|\nabla(B^{2}/8\pi)|}{\rho\Omega^{2}r},
\end{equation}
where $\Omega=2\pi\nu_{\phi}$ is the Keplerian angular velocity.
Based on equation (10) and the values of the parameters given in
Table 1 we have the ratios versus disc radius as shown in Figure 3,
from which we find that the ratios are much less than unity for
these sources. Thus the ``weak magnetic field assumption" remains
effective in our model.

\subsection{Estimating the BH Mass of H1743-322 based on 3:2 HFQPO pair}

Since HFQPOs are probably produced in the inner disc region very
close to ISCO, they might offer the most reliable measurement of BH
spins (MR06; RM06). Compared with other methods for measuring spin,
such as fitting the spectrum of the X-ray continuum, HFQPO method is
much more simple, being independent of disc inclination relative to
the BH's spin axis. Furthermore, the 3:2 HFQPO pairs provide a very
strict constraint to the BH spin, provided the BH mass is estimated.

A very narrow range of the BH spin, $0.989<a_{*}<0.994$, can be
found in Table 2, which corresponds to the upper limit to the BH
mass for fitting the 3:2 HFQPO pairs associated with the SPL states
observed in the three BH binaries. Combining this narrow range of
the BH spin with equations (5)---(7), we obtain a very simple
relation between BH mass and the upper frequency of the 3:2 HFQPO
pairs as follows,

\begin{equation}
   m_{\textup{\scriptsize BH}}\nu_{\textup{\scriptsize up}}=3000,
\end{equation}
where the upper frequency $\nu_{\textup{\scriptsize
up}}=\nu_{\theta}$ is equal to the vertical resonance frequency
given in equation (6). It is noted that $m_{\textup{\scriptsize
BH}}$ in equation (11) is the upper limit to the BH mass. Replacing
$\nu_{\textup{\scriptsize up}}$ in equation (11) by $3\nu_{0}$, we
have almost the same relation
$\nu_{0}(\textup{Hz})=931m_{\textup{\scriptsize BH}}^{-1}$ given in
MR06.

The 3:2 HFQPO pair has been observed in the bright X-ray transient
H1743-322. Although the mass of its BH primary has not been
measured, its behavior resembles the BH binaries XTE J1550-564 and
GRO J1655-40 in many ways \citep{b8,b9,b23,b26}. Very recently,
\citet{b14} pointed out that H1743-322 does contain a BH primary
based on a detailed analysis of its strong similarities to XTE
J1550-564. As a simple analysis, we have the upper limit to the BH
mass of H1743-322, $m_{\textup{\scriptsize BH}}<12.5$, by
substituting $\nu_{\textup{\scriptsize up}}=240$Hz into equation
(11). We expect that this upper limit can be checked by future
observations on H1743-322.

\section*{Acknowledgments}

This work is supported by the National Natural Science Foundation of
China under grant 10873005, the Research Fund for the Doctoral
Program of Higher Education under grant 200804870050 and National
Basic Research Program of China under grant 2009CB824800. We are
very grateful to the anonymous referee for his (her) helpful
suggestion for improving our work.

\label{lastpage}

\end{document}